\documentclass[twocolumn,showpacs,preprintnumbers, superscriptaddress]{revtex4}
\usepackage{graphicx}
\usepackage{dcolumn}
\usepackage{bm}

\begin{document}

\title{Two-band model of Raman scattering on electron-doped high-$T_c$
superconductor}
\author{C. S. Liu}
\affiliation{Institute of Theoretical Physics and Interdisciplinary Center of Theoretical
Studies, Chinese Academy of Sciences, P. O. Box 2735, Beijing 100080, China}
\affiliation{Department of Physics, National Taiwan Normal University, Taipei 11650,
Taiwan}
\author{H. G. Luo}
\affiliation{Institute of Theoretical Physics and Interdisciplinary Center of Theoretical
Studies, Chinese Academy of Sciences, P. O. Box 2735, Beijing 100080, China}
\author{W. C. Wu}
\affiliation{Department of Physics, National Taiwan Normal University, Taipei 11650,
Taiwan}
\author{T. Xiang}
\affiliation{Institute of Theoretical Physics and Interdisciplinary Center of Theoretical
Studies, Chinese Academy of Sciences, P. O. Box 2735, Beijing 100080, China}
\affiliation{Center for Advanced Study, Tsinghua University, Beijing 100084, China}

\begin{abstract}
We have analyzed the $B_{1g}$ and $B_{2g}$ Raman spectra of electron-doped
cuprate superconductors Nd$_{2-x}$Ce$_{x}$CuO$_4$ and Pr$_{2-x}$Ce$_{x}$CuO$%
_4$ using a weakly coupled two-band model. One of these two bands is
centered around $(\pm \pi/2, \pm \pi/2)$ and couples more strongly with the $%
B_{2g}$ mode, while the other is centered around $(\pm \pi , 0)$ and $(0,
\pm \pi)$ and couples more strongly with the $B_{1g}$ mode. This model
explains in a natural way why the $B_{2g}$ Raman peak occurs at a higher
frequency than the $B_{1g}$ one at optimal doping, and how these two peaks
change with doping in agreement with experiments. The result thus supports
that there are two kinds of quasiparticles in electron-doped cuprates and $%
d_{x^2-y^2}$-wave superconductivity is driven by the holelike band and a
proximity effect on the electronlike band.
\end{abstract}

\maketitle

\section{Introduction}

Pairing symmetry of electron-doped high-$T_c$ cuprate superconductors such
as Nd$_{2-x}$Ce$_{x}$CuO$_4$ and Pr$_{2-x}$Ce$_{x}$CuO$_4$ is a long
standing problem \cite{tsuei2000, armitage2001, sato2001, alff1999,
prozorov2000, skinta2002, kim2003, biswas2002, chesca2003, chesca2005,
shan2005, balci2004, stadlober1995, blumberg2002, qazilbash2005,
matsui2005a, ariando2005}. Although no consensus has been reached yet, more
and more recent experimental results have suggested that the order parameter
of electron-doped cuprates is likely to have $d_{x^{2}-y^{2}}$-wave pairing
symmetry \cite{tsuei2000, armitage2001, sato2001, prozorov2000,
blumberg2002, chesca2003, chesca2005, matsui2005a, ariando2005}, in close
resemblance to that of hole-doped materials. Among various experiments, the
Raman scattering seems to suggest a different story \cite{stadlober1995,
blumberg2002, qazilbash2005}. For hole-doped superconductors, it was known
that typical $B_{1g}$, $B_{2g}$, and $A_{1g}$ pair-breaking peaks appear,
respectively, at the frequencies of $2$, $1.6$, and $1.2$ times of the gap
amplitude \cite{devereaux1995}. However, in electron-doped materials the
relative position of the $B_{1g}$ and $B_{2g}$ peaks changes with doping.
The $B_{2g}$ peak appears first at a higher frequency than the $B_{1g}$ one
in the underdoped regime. It then moves down and finally appears at a
frequency lower than that of the $B_{1g}$ peak in the heavily overdoped
regime.

The Raman scattering has the potential to probe different regions of the
Fermi surface (FS), thus a thorough understanding of the experimental data
can provide a better understanding on the momentum dependence of
superconducting (SC) pairing gap. The observation of $B_{2g}$ Raman peak at
higher frequency than that of $B_{1g}$ would imply a non-monotonic $%
d_{x^2-y^2}$-wave order parameter in a single-band system \cite{blumberg2002}%
. This nonmonotonic order parameter seems to be also consistent with the
observation of angle resolved photoemission spectroscopy (ARPES) \cite%
{matsui2005a}. However, this one-band picture may not be adequate to
describe the nature of two kinds of charge carriers in electron-doped
cuprate superconductors, as revealed by magneto-transport measurements \cite%
{wang1991, jiang1994, fournier1997}.

A key clue towards the understanding of Raman data in Nd$_{2-x}$Ce$_{x}$CuO$%
_4$ comes from the doping evolution of the FS revealed by ARPES \cite%
{armitage2002, matsui2005b}. At low doping, four small FS pockets first
appear around $(\pm\pi ,0)$ and $(0,\pm\pi )$. By increasing the doping,
four new pockets begin to form around $(\pm\pi /2,\pm\pi /2)$. These results
can be explained in terms of the \textbf{k}-dependent band-folding effect
due to antiferromangetic (AF) ordering. The original band is folded back
into the magnetic Brillouin zone (MBZ) around the diagonal line ($\pi,
0)\rightarrow(0,\pi$). Near the intersecting points of the Fermi surface, an
AF gap opens and splits the original FS into two \cite{kusko2002, yuan2004,
voo2005}. This two-band picture was first used by Luo and Xiang to explain
the unusual temperature dependence of the magnetic penetration depth in
electron-doped copper oxides \cite{luo2005}. It is supported by Hall
coefficient and magneto-resistance measurements \cite{wang1991, jiang1994,
fournier1997}. The generic feature of a weakly coupled two-band model was
discussed in Ref. \cite{Xiang1996} in the context of hole-doped cuprate
superconductors.

In this paper, we shall use a two-band model to study the Raman response for
the electron-doped cuprates. As will be shown, the two-band model gives a
unified explanation to the unusual behaviors of Raman spectra. It explains
in a natural way why the $B_{2g}$ Raman peak appears at a higher frequency
than that of the $B_{1g}$ peak at optimal doping and why the relative
positions of these two peaks change in the heavily overdoped regime.

\section{Model and Formalism}

We start by considering the two-dimensional $t$-$t^{\prime }$-$t^{\prime
\prime }$-$J$ model
\begin{eqnarray}
H &=&-t\sum_{\langle ij\rangle _{1}\sigma }c_{i\sigma }^{\dagger }c_{j\sigma
}-t^{\prime }\sum_{\langle ij\rangle _{2}\sigma }c_{i\sigma }^{\dagger
}c_{j\sigma }-t^{\prime \prime }\sum_{\langle ij\rangle _{3}\sigma
}c_{i\sigma }^{\dagger }c_{j\sigma }  \nonumber \\
&&+J\sum_{\langle ij\rangle }\left( \vec{S}_{i}\cdot \vec{S}_{j}-{\frac{1}{4}%
}n_{i}n_{j}\right) ,  \label{initial Hamiltonian}
\end{eqnarray}%
where $\langle ij\rangle _{1}$, $\langle ij\rangle _{2}$, and $\langle
ij\rangle _{3}$ denote the nearest, second-nearest, and third-nearest
neighbors between $i$ and $j$. No double occupied sites are allowed in Eq.
(1). All notations used in (\ref{initial Hamiltonian}) are standard.
Applying the slave-boson and MF decoupling \cite{yuan2004}, Hamiltonian (\ref%
{initial Hamiltonian}) can be written in terms of two (diagonalized) bands
in momentum space
\[
H={\sum_{\mathbf{k}\sigma }}^{\prime }(\xi _{\mathbf{k}\alpha }\alpha _{%
\mathbf{k}\sigma }^{\dagger }\alpha _{\mathbf{k}\sigma }+\xi _{\mathbf{k}%
\beta }\beta _{\mathbf{k}\sigma }^{\dagger }\beta _{\mathbf{k}\sigma }),
\]%
where the prime denotes that the momentum summation is over the MBZ only ($%
-\pi <k_{x}\pm k_{y}\leq \pi $) and
\[
\xi _{\mathbf{k},l}=\frac{\varepsilon _{\mathbf{k}}+\varepsilon _{\mathbf{k+Q%
}}}{2}\mp \sqrt{\frac{(\varepsilon _{\mathbf{k+Q}}-\varepsilon _{\mathbf{k}%
})^{2}}{4}+4J^{2}m^{2}}
\]%
with $\mathbf{Q}\equiv (\pi ,\pi )$ the AF wave vector and $m\equiv
(-1)^{i}\langle S_{i}^{z}\rangle $ the AF order. Here
\begin{eqnarray}
\varepsilon _{\mathbf{k}} &=&(2|t|\delta -J\chi )\left( \cos k_{x}+\cos
k_{y}\right)   \nonumber \\
&-&4t^{\prime }\delta \cos k_{x}\cos k_{y}-2t^{\prime \prime }\delta (\cos
2k_{x}+\cos 2k_{y})  \label{energy
band dispersion}
\end{eqnarray}%
with $\chi \equiv \langle f_{i\sigma }^{\dag }f_{j\sigma }\rangle $ the
uniform bond order and $\delta $ the doping concentration ($f_{i\sigma }$ is
the fermionic spinon operator).

In the SC state, we add a BCS coupling term to each band and assume the
system to be described by the following Hamiltonian \cite{luo2005}
\[
H={\sum_{\mathbf{k\sigma }l}}^{\prime }\xi _{\mathbf{k}l}l_{\mathbf{k}\sigma
}^{\dagger }l_{\mathbf{k}\sigma }+{\sum_{\mathbf{k}l}}^{\prime }\Delta _{%
\mathbf{k}l}(l_{\mathbf{k}\uparrow }^{\dag }l_{-\mathbf{k}\downarrow }^{\dag
}+l_{-\mathbf{k}\downarrow }l_{\mathbf{k}\uparrow }),
\]%
where $l\equiv \alpha ,\beta $ and $\Delta _{\mathbf{k},l}=\left( \Delta
_{l}/2\right) [\cos k_{x}-\cos k_{y}]$ are the $d_{x^{2}-y^{2}}$-wave gap
functions.

The Raman scattering intensity is proportional to the imaginary part of the
effective density-density correlation function $\chi (\mathbf{q},\tau
)=\langle T_\tau \lbrack \tilde{\rho}( \mathbf{q},\tau),\tilde{\rho}(-%
\mathbf{q},0)]\rangle$ in the limit $\mathbf{q}\rightarrow 0$. Here $\tilde{%
\rho}(\mathbf{q},\tau ) \equiv {\sum_{\mathbf{k},\sigma }}\gamma _{\mathbf{k}%
}c_{\mathbf{k}+\mathbf{q}}^{\dag }(\tau )c_{\mathbf{k}}(\tau)$ is the
effective density operator and $\gamma _{\mathbf{k}}$ is the Raman vertex.
When the energy of incident light is smaller than the optical band gap, the
contribution from the resonance channel is negligible. The Raman vertex can
then be obtained in terms of the curvature of the band dispersion under the
inverse effective mass approximation.

In the current two-band model, the effective density operator is decomposed
as
\begin{eqnarray}
\tilde{\rho}(\mathbf{q},\tau ) &\equiv &{\sum_{\mathbf{k},\sigma }}^{\prime
}[\gamma _{\mathbf{k}}c_{\mathbf{k}+\mathbf{q},\sigma }^{\dag }(\tau )c_{%
\mathbf{k},\sigma }(\tau )  \nonumber \\
&+&\gamma _{\mathbf{k+Q}}c_{\mathbf{k+Q}+\mathbf{q},\sigma }^{\dag }(\tau
)c_{\mathbf{k+Q},\sigma }(\tau )].
\end{eqnarray}%
Along with the unitary transformation \cite{yuan2004} such that operators $%
c_{\mathbf{k}}$ and $c_{\mathbf{k}+\mathbf{Q}}$ are transformed into $\alpha
_{\mathbf{k}}$ and $\beta _{\mathbf{k}+\mathbf{Q}}$, the Raman response
function for each symmetric channel ($\mathrm{S}$) is then given by \cite%
{csliu}
\begin{eqnarray}
\chi ^{S}(\mathbf{q} &\rightarrow &0,\tau )=-{\sum_{\mathbf{k,}ll^{\prime }}}%
^{\prime }(\gamma _{\mathbf{k},ll^{\prime }}^{S})^{2}[\mathcal{G}_{l}(%
\mathbf{k},\tau )\mathcal{G}_{l^{^{\prime }}}(\mathbf{k},-\tau )  \nonumber
\\
&&-\epsilon _{ll^{\prime }}\mathcal{F}_{l}(\mathbf{k},\tau )\mathcal{F}%
_{l^{^{\prime }}}(\mathbf{k},-\tau )],  \label{response
function}
\end{eqnarray}%
where $\mathcal{G}$ and $\mathcal{F}$ are the normal and anomalous Green
functions for a superconductor, $\epsilon _{ll^{\prime }}=1$ if $l=l^{\prime
}$ or $-1$ if $l\neq l^{\prime }$. The intra- and interband vertex functions
are
\begin{eqnarray}
\gamma _{\mathbf{k},\alpha \alpha }^{S} &=&\cos ^{2}\theta _{\mathbf{k}%
}\gamma _{\mathbf{k}}^{S}+\sin ^{2}\theta _{\mathbf{k}}\gamma _{\mathbf{k+Q}%
}^{S},  \nonumber \\
\gamma _{\mathbf{k},\beta \beta }^{S} &=&\sin ^{2}\theta _{\mathbf{k}}\gamma
_{\mathbf{k}}^{S}+\cos ^{2}\theta _{\mathbf{k}}\gamma _{\mathbf{k+Q}}^{S},
\label{vertex functions} \\
\gamma _{\mathbf{k},\alpha \beta }^{S} &=&\gamma _{\mathbf{k},\beta \alpha
}^{S}=\sin 2\theta _{\mathbf{k}}(\gamma _{\mathbf{k}}^{S}-\gamma _{\mathbf{%
k+Q}}^{S}),  \nonumber
\end{eqnarray}%
where $\cos 2\theta _{\mathbf{k}}\equiv (\varepsilon _{\mathbf{k+Q}%
}-\varepsilon _{\mathbf{k}})/\sqrt{(\varepsilon _{\mathbf{k+Q}}-\varepsilon
_{\mathbf{k}})^{2}+4J^{2}m^{2}}$. For the $B_{1g}$ and $B_{2g}$ channels, $%
\gamma _{\mathbf{k}}^{B_{1g}}=\gamma _{\mathbf{k}}^{xx}-\gamma _{\mathbf{k}%
}^{yy}$ and $\gamma _{\mathbf{k}}^{B_{2g}}=2\gamma _{\mathbf{k}}^{xy}$. Here
$\gamma _{\mathbf{k}}^{ij}\equiv \partial ^{2}\varepsilon _{\mathbf{k}%
}/\partial k_{i}\partial k_{j}$ (inverse effective mass approximation). Eq.~(%
\ref{response function}) and (\ref{vertex functions}) reduce to the famous
ones in a one-band system when $m=0$ \cite{devereaux1995}.

\begin{figure}[htbp]
\includegraphics[width=7cm]{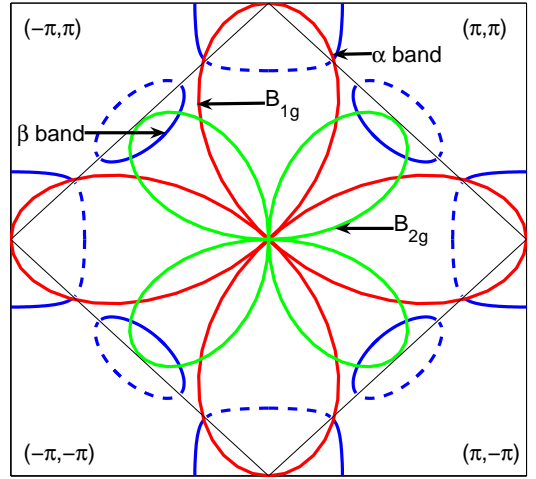}
\caption{Schematic plot of the Fermi surfaces for the $\protect\alpha$ and $%
\protect\beta$ bands and the momentum dependence of the $B_{1g}$ and $B_{2g}$
Raman vertices. The $B_{1g}$ and $B_{2g}$ modes couple more strongly with
the $\protect\alpha$ and $\protect\beta$ bands, respectively.}
\label{fig1}
\end{figure}

Shown in Fig.~\ref{fig1} are the $B_{2g}$ and $B_{1g}$ Raman vertices and
how they are coupled to the SC quasiparticle excitations in \textbf{k}
space. Since the $B_{2g}$ vertex has $d_{xy}$ symmetry, the $B_{2g}$ channel
is dominated by the excitations of the $\beta$ band. In contrast, the $B_{1g}
$ vertex has $d_{x^{2}-y^{2}}$ symmetry, thus the $B_{1g}$ channel is
contributed mainly from the excitations of the $\alpha$ band. Since both $%
B_{2g}$ and $B_{1g}$ vertices are odd-parity, their Raman intensities are
not affected by the Coulomb screening. For the fully symmetric $A_{1g}$
channel, in contrast, all regions of momentum space contribute and the Raman
intensity is partially screened. Since the $A_{1g}$ channel is more
sensitive to the actual vertex as well as the screening effect, we will
leave out $A_{1g}$ and focus on the $B_{2g}$ and $B_{1g}$ channels only.

In electron doped cuprates, the SC state appears only when the $\beta $ band
emerges above the Fermi energy. This would suggest that we assume that it is
the $\beta $ band that drives the system to superconduct, while the $\alpha $
band becomes superconducting mainly via the proximity effect. A simple
picture for the understanding of the unusual Raman spectra in electron doped
cuprates can then be sketched as follows. In the under doped or optimally
doped regime, the $\beta $ band couples more strongly with the AF
fluctuations than the $\alpha $ band. This results in a relatively larger SC
gap in the $\beta $ band ($\Delta _{\beta }$) than in the $\alpha $ band ($%
\Delta _{\alpha }$). The $B_{1g}$ channel probes mainly the quasi-particle
(QP) excitations of the $\alpha $ band, thus the $B_{1g}$ Raman peak is
mainly determined by $\Delta _{\alpha }$. Similarly, the $B_{2g}$ channel
probes mainly the QP excitations of the $\beta $ band, its Raman peak is
mainly determined by $\Delta _{\beta }$. If $\Delta _{\alpha }$ is much
smaller than $\Delta _{\beta }$, one would then expect that the $B_{1g}$
peak appears at a frequency lower than that of the $B_{2g}$ peak, unlike in
the hole doped case. In the heavily overdoped regime, the AF correlation
becomes very weak and the band splitting vanishes. In this case the two-band
model reduces essentially to a one-band model and the $B_{1g}$ and $B_{2g}$
Raman spectra would behave similarly as in the hole doped cuprate
superconductors, consistent with the experiments.

\section{Results and Discussions}

\subsection{On Nd$_{2-x}$Ce$_{x}$CuO$_4$}

Pertaining to Nd$_{2-x}$Ce$_{x}$CuO$_{4}$, we have adopted $|t|=0.326$ $%
\mathrm{eV}$, $t^{\prime }=0.3t$, $t^{\prime \prime }=-0.2t$, and $J=0.3t$
to simulate the band structure \cite{yuan2004}. We take $\chi =-0.15$ and $%
m=0.178$ for the optimally doped ($x=0.15$) sample, $\chi =-0.15$ and $m=0.15
$ for the overdoped ($x=0.16$) sample. These parameters are close to those
obtained in self-consistent calculations for the normal states \cite%
{yuan2004}. The chemical potentials are determined by the filling factor for
each band to give the true doping concentration through $x=n_{e}-n_{h}$.

Theoretical fitting procedures are implemented as follows. First, the vertex
functions are evaluated using (\ref{vertex functions}). Then the SC gaps $%
\Delta _{\alpha }$ and $\Delta _{\beta }$ and the smearing Lorentz width $%
\Gamma _{\alpha }$ and $\Gamma _{\beta }$ are adjusted to fit the peak
positions and the overall spectral line shape (up to a constant multiplying
factor).

\begin{figure}[htbp]
\includegraphics[width = 8.5cm]{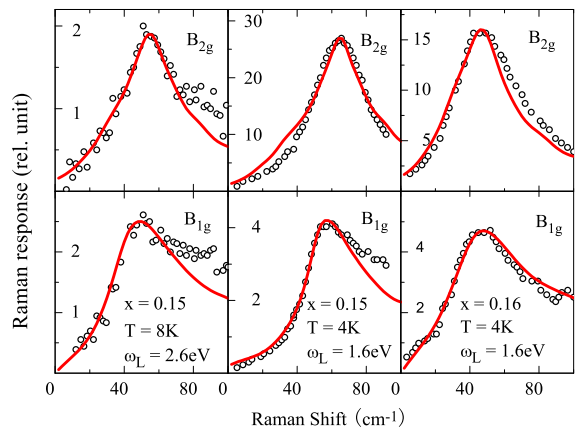}
\caption{Comparison of the theoretical calculations (solid lines) with the
measurement data (circles) for Nd$_{2-x}$Ce$_{x}$CuO$_4$. The experimental
data in the first column are taken from Ref.~\protect\cite{blumberg2002}.
The data in the second and third columns are taken from Ref.\protect\cite%
{qazilbash2005}. $\protect\omega_L$ is the incident photon energy. }
\label{fig2}
\end{figure}

Figure~\ref{fig2} compares the experimental data of Raman spectra for Nd$%
_{2-x}$Ce$_{x}$CuO$_{4}$ with the theoretical calculations. For all the
cases considered in Fig.~\ref{fig2}, our fitting curves are in good
agreement with the experimental results. For the optimally doped sample
(x=0.15) reported in Ref. \cite{blumberg2002} at $T=8$ $K$ (first column),
the $B_{1g}$ and $B_{2g}$ Raman peaks appear at $50$ $\mathrm{cm}^{-1}$ and $%
55$ $\mathrm{cm}^{-1}$, respectively. The corresponding gap and smearing
parameters obtained by fitting are $(\Delta _{\alpha },\Delta _{\beta })=(21$
$\mathrm{cm}^{-1},48$ $\mathrm{cm}^{-1})$ and $(\Gamma _{\alpha },\Gamma
_{\beta })=(6$ $\mathrm{cm}^{-1},8$ $\mathrm{cm}^{-1}$). The ratio between
the $B_{1g}$ Raman peak frequency and $\Delta _{\alpha }$ is about $2.4$,
while the ratio between the $B_{2g}$ Raman peak frequency and $\Delta
_{\beta }$ is about $1.2$. The corresponding ratios in a hole doped d$%
_{x^{2}-y^{2}}$-wave superconductor are about 2 and 1.6, respectively. This
difference between hole and electron doped cuprates is not difficult to be
understood. In electron doped materials, the AF correlation splits the
continuous FS into two separate sheets. This then suppresses the high (low)
energy region to which the $B_{2g}$ ($B_{1g}$) probes. It is thus expected
that the $B_{2g}$ ($B_{1g}$) peak will be red shifted (blue shifted)
compared with the result of hole-doped cuprate superconductors.

For another set of data of the optimally doped sample reported in Ref.~\cite%
{qazilbash2005} at $T=4\mathrm{k}$ (second column), the fitting parameters
are $(\Delta _{\alpha },\Delta _{\beta })=(27$ $\mathrm{cm}^{-1},57$ $%
\mathrm{cm}^{-1})$ and $(\Gamma _{\alpha },\Gamma _{\beta })=(4$ $\mathrm{cm}%
^{-1},6$ $\mathrm{cm}^{-1})$. The corresponding parameters for the overdoped
sample at $T=4$ $K$ (third column) are $(\Delta _{\alpha },\Delta _{\beta
})=(21$ $\mathrm{cm}^{-1},37$ $\mathrm{cm}^{-1})$ and $\Gamma _{\alpha
}=\Gamma _{\beta }=8$ $\mathrm{cm}^{-1}$. The gap parameters obtained are
consistent with the general expectation. Both $\Delta _{\alpha }$ and $%
\Delta _{\beta }$ decrease with increasing temperature at the same doping
level and with increasing doping at the same temperature. The gap ratio, $%
r\equiv \Delta _{\beta }/\Delta _{\alpha }$, is reduced from 2 at optimal
doping to 1.7 at slightly overdoping, consistent with the scenario of
AF-like fluctuation induced superconductivity. All fitting parameters for Nd$%
_{2-x}$Ce$_{x}$CuO$_{4}$~ are summarized in Table~\ref{NdCe}.

\begin{table}[tbp]\caption{Summary of fitting parameters for Nd$_{2-x}$Ce$_{x}$CuO$_4$~
(refer to Fig.~2).}
\begin{ruledtabular}
\begin{tabular}{cccc}
$ $ & Column 1 & Column 2&  Column 3
\\ \hline $T_c$ (K) & 22 & 22 &  13\\
 $m$ & 0.178 & 0.178 &  0.15\\
$\Gamma_\alpha$$(\mathrm{cm}^{-1})$ & 6 & 4 &  8\\
$\Gamma_\beta$$(\mathrm{cm}^{-1})$ & 8 & 6 &  8\\
$\Delta_\alpha$$(\mathrm{cm}^{-1})$ & 21 & 27 & 21 \\
$\Delta_\beta$$(\mathrm{cm}^{-1})$ & 48 & 57 & 37
\\$\Delta_\beta/\Delta_\alpha$ &
2.29 & 2.11 & 1.76 \\
$\Delta_\beta/T_c$ & 2.18 & 2.59 & 2.84 \\
\end{tabular}
\end{ruledtabular}
\label{NdCe}
\end{table}

\begin{table}[tbp]
\caption{Summary of fitting parameters for
Pr$_{2-x}$Ce$_{x}$CuO$_4$~ (refer to Fig.~3).} \label{PrCe}
\begin{ruledtabular}
\begin{tabular}{cccc}
 & Column 1 &  Column 2 & Column 3
\\ \hline $x$ & 0.15 & 0.165 &  0.18
\\ $T_c$ (K) & 23.5 & 15 & 10 \\
 $m$ & 0.15 & 0.12 &   0 \\
$\Gamma_\alpha$$(\mathrm{cm}^{-1})$ & 6 & 8 &  15\\
$\Gamma_\beta$$(\mathrm{cm}^{-1})$ & 6 & 8 &  15\\
$\Delta_\alpha$$(\mathrm{cm}^{-1})$ & 31 & 16 & 15 \\
$\Delta_\beta$$(\mathrm{cm}^{-1})$ &
68 & 30 & 15 \\
$\Delta_\beta/\Delta_\alpha$ &
2.19 & 1.88 & 1 \\
$\Delta_\beta/T_c$ & 2.89 & 2 & 1.5 \\
\end{tabular}
\end{ruledtabular}
\end{table}

\subsection{On Pr$_{2-x}$Ce$_{x}$CuO$_4$}

\begin{figure}[htbp]
\includegraphics[width=8.5cm]{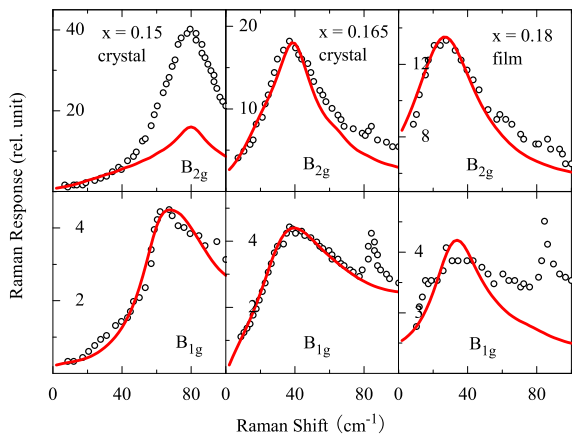}
\caption{ Comparison between the theoretical results (solid curves) and the
Raman measurement data from Ref.~\protect\cite{qazilbash2005} (circles) for
Pr$_{2-x}$Ce$_{x}$CuO$_4$ at $T=4 \mathrm{K}$. $\protect\omega_L=1.9 \mathrm{%
eV}$. }
\label{fig3}
\end{figure}

The Raman scattering measurement has also been carried out in electron-doped
Pr$_{2-x}$Ce$_{x}$CuO$_4$ at various doping levels \cite{qazilbash2005}. The
Raman spectra of Pr$_{2-x}$Ce$_{x}$CuO$_4$, as shown in Fig.~\ref{fig3},
behave similarly as for Nd$_{2-x}$Ce$_{x}$CuO$_4$. At optimal doping, the $%
B_{2g}$ peak appears at a frequency higher than that of the $B_{1g}$ peak.
With increasing doping, the frequency of the $B_{1g}$ peak approaches to and
finally surpasses the $B_{2g}$ peak in the overdoped regime.

Figure.~\ref{fig3} compares the measurement data of Pr$_{2-x}$ Ce$_{x}$CuO$%
_{4}$ with our theoretical calculations. For the optimally doped sample ($%
x=0.15$), the Raman peak appears at 80 $\mathrm{cm}^{-1}$ for the
$B_{2g}$ mode and at 62 $\mathrm{cm}^{-1}$ for the $B_{1g}$ mode.
The parameters obtained by fitting are $(\Delta _{\alpha },\Delta
_{\beta })=(31$ $\mathrm{cm}^{-1},68$ $\mathrm{cm}^{-1})$ and
$\Gamma _{\alpha }=\Gamma _{\beta }=6$ $\mathrm{cm}^{-1}$, with
$\chi =-0.15$ and $m=0.15$. For the slightly overdoped sample
($x=0.165$),
the $B_{2g}$ and $B_{1g}$ peaks appear at the same frequency at 37 $\mathrm{%
cm}^{-1}$ and the parameters we obtained are $(\Delta _{\alpha
},\Delta _{\beta })=(16$ $\mathrm{cm}^{-1},30$ $\mathrm{cm}^{-1})$
and $\Gamma _{\alpha }=\Gamma _{\beta }=8$ $\mathrm{cm}^{-1}$, with
$\chi =-0.15$ and $m=0.12$. For the heavily over-doped sample
($x=0.18$), the Raman peaks appear at 25 $\mathrm{cm}^{-1}$ and 30
$\mathrm{cm}^{-1}$ for the $B_{2g}$ and $B_{1g}$ modes,
respectively.
The relative peak positions of these two modes are similar as in a one-band $%
d_{x^{2}-y^{2}}$-wave superconductor. This is not unexpected since
at such a high doping level, the two-band model reduces essentially
to a one-band model. In this case, the parameters we obtained are
$\Delta _{\alpha }=\Delta _{\beta }=15$ $\mathrm{cm}^{-1}$ and
$\Gamma _{\alpha }=\Gamma _{\beta }=15$ $\mathrm{cm}^{-1}$, with
$\chi =-0.15$ and $m=0$. For the above three samples, the ratio
$\Delta _{\beta }/\Delta _{\alpha }$ changes from $2.3$, to $1.9$,
and finally to $1$ with increasing doping. With increasing doping,
the AF order is depressed and the gap amplitudes is decreased. The
results are consistent with neutron scattering measurement
\cite{yamada137004}. All fitting parameters for Pr$_{2-x}$Ce$_{x}$%
CuO$_4$~ are summarized in TABLE~\ref{PrCe}.

In the top-left panel of Fig.~\ref{fig3} for the $B_{2g}$ mode, the
experimental data are taken under the strong \emph{resonant} regime
as emphasized in Refs. \cite{blumberg2002, qazilbash2005}. In this
case, the contribution from the resonance channel becomes important
and the inverse effective mass approximation is not valid. Our
theoretical result including only the non-resonant contribution can
give a good explanation to the low-frequency part of the spectrum,
but the height of the peak is much lower than the experimental one.
The resonance channel may also have some contribution to the
$B_{2g}$ spectrum in the second column of Fig.~\ref{fig2} for
Nd$_{2-x}$ Ce$_{x}$CuO$_4$.

\section{Summary}

In summary, we have analyzed the Raman spectra of electron-doped cuprate
superconductors based on a weakly coupled two-band model. Our result gives a
unified explanation to the experimental data in the whole doping range. It
suggests strongly that the SC pairing in electron-doped cuprate
superconductors results from the same pairing mechanism as in hole doped
ones and the gap parameter has $d_{x^{2}-y^{2}}$-wave symmetry. To
understand the Raman data in the strong resonance regime, a more
comprehensive theory including the contribution from the resonance channel
is desired.

\begin{acknowledgements}
The authors are grateful to Hsiang-Lin Liu for useful comments on
Raman experiments. This work was supported by the National Natural
Science Foundation of China (Grant No. 10347149), National Basic
Research Program of China (Grant No. 2005CB32170X), and National
Science Council of Taiwan (Grant No. 93-2112-M-003-015).
\end{acknowledgements}

\bibliography{Raman_PRB}

\begin{thebibliography}{30}
\expandafter\ifx\csname natexlab\endcsname\relax\def\natexlab#1{#1}\fi
\expandafter\ifx\csname bibnamefont\endcsname\relax
  \def\bibnamefont#1{#1}\fi
\expandafter\ifx\csname bibfnamefont\endcsname\relax
  \def\bibfnamefont#1{#1}\fi
\expandafter\ifx\csname citenamefont\endcsname\relax
  \def\citenamefont#1{#1}\fi
\expandafter\ifx\csname url\endcsname\relax
  \def\url#1{\texttt{#1}}\fi
\expandafter\ifx\csname urlprefix\endcsname\relax\def\urlprefix{URL }\fi
\providecommand{\bibinfo}[2]{#2}
\providecommand{\eprint}[2][]{\url{#2}}

\bibitem[{\citenamefont{Tsuei and Kirtley}(2000)}]{tsuei2000}
\bibinfo{author}{\bibfnamefont{C.~C.} \bibnamefont{Tsuei}} \bibnamefont{and}
  \bibinfo{author}{\bibfnamefont{J.~R.} \bibnamefont{Kirtley}},
  \bibinfo{journal}{Phys. Rev. Lett.} \textbf{\bibinfo{volume}{85}},
  \bibinfo{pages}{182} (\bibinfo{year}{2000}).

\bibitem[{\citenamefont{Armitage et~al.}(2001)\citenamefont{Armitage, Lu, Feng,
  Kim, Damascelli, Shen, Ronning, Shen, Onose, Taguchi et~al.}}]{armitage2001}
\bibinfo{author}{\bibfnamefont{N.~P.} \bibnamefont{Armitage}},
  \bibinfo{author}{\bibfnamefont{D.~H.} \bibnamefont{Lu}},
  \bibinfo{author}{\bibfnamefont{D.~L.} \bibnamefont{Feng}},
  \bibinfo{author}{\bibfnamefont{C.}~\bibnamefont{Kim}},
  \bibinfo{author}{\bibfnamefont{A.}~\bibnamefont{Damascelli}},
  \bibinfo{author}{\bibfnamefont{K.~M.} \bibnamefont{Shen}},
  \bibinfo{author}{\bibfnamefont{F.}~\bibnamefont{Ronning}},
  \bibinfo{author}{\bibfnamefont{Z.-X.} \bibnamefont{Shen}},
  \bibinfo{author}{\bibfnamefont{Y.}~\bibnamefont{Onose}},
  \bibinfo{author}{\bibfnamefont{Y.}~\bibnamefont{Taguchi}},
  \bibnamefont{et~al.}, \bibinfo{journal}{Phys. Rev. Lett.}
  \textbf{\bibinfo{volume}{86}}, \bibinfo{pages}{1126} (\bibinfo{year}{2001}).

\bibitem[{\citenamefont{Sato et~al.}(2001)\citenamefont{Sato, Kamiyama,
  Takahashi, Kurahashi, and Yamada}}]{sato2001}
\bibinfo{author}{\bibfnamefont{T.}~\bibnamefont{Sato}},
  \bibinfo{author}{\bibfnamefont{T.}~\bibnamefont{Kamiyama}},
  \bibinfo{author}{\bibfnamefont{T.}~\bibnamefont{Takahashi}},
  \bibinfo{author}{\bibfnamefont{K.}~\bibnamefont{Kurahashi}},
  \bibnamefont{and} \bibinfo{author}{\bibfnamefont{K.}~\bibnamefont{Yamada}},
  \bibinfo{journal}{Science} \textbf{\bibinfo{volume}{291}},
  \bibinfo{pages}{1517} (\bibinfo{year}{2001}).

\bibitem[{\citenamefont{Alff et~al.}(1999)\citenamefont{Alff, Meyer, Kleefisch,
  Schoop, Marx, Sato, Naito, and Gross}}]{alff1999}
\bibinfo{author}{\bibfnamefont{L.}~\bibnamefont{Alff}},
  \bibinfo{author}{\bibfnamefont{S.}~\bibnamefont{Meyer}},
  \bibinfo{author}{\bibfnamefont{S.}~\bibnamefont{Kleefisch}},
  \bibinfo{author}{\bibfnamefont{U.}~\bibnamefont{Schoop}},
  \bibinfo{author}{\bibfnamefont{A.}~\bibnamefont{Marx}},
  \bibinfo{author}{\bibfnamefont{H.}~\bibnamefont{Sato}},
  \bibinfo{author}{\bibfnamefont{M.}~\bibnamefont{Naito}}, \bibnamefont{and}
  \bibinfo{author}{\bibfnamefont{R.}~\bibnamefont{Gross}},
  \bibinfo{journal}{Phys. Rev. Lett.} \textbf{\bibinfo{volume}{83}},
  \bibinfo{pages}{2644} (\bibinfo{year}{1999}).

\bibitem[{\citenamefont{Prozorov et~al.}(2000)\citenamefont{Prozorov,
  Giannetta, Fournier, and Greene}}]{prozorov2000}
\bibinfo{author}{\bibfnamefont{R.}~\bibnamefont{Prozorov}},
  \bibinfo{author}{\bibfnamefont{R.~W.} \bibnamefont{Giannetta}},
  \bibinfo{author}{\bibfnamefont{P.}~\bibnamefont{Fournier}}, \bibnamefont{and}
  \bibinfo{author}{\bibfnamefont{R.~L.} \bibnamefont{Greene}},
  \bibinfo{journal}{Phys. Rev. Lett.} \textbf{\bibinfo{volume}{85}},
  \bibinfo{pages}{3700} (\bibinfo{year}{2000}).

\bibitem[{\citenamefont{Skinta et~al.}(2002)\citenamefont{Skinta, Kim,
  Lemberger, Greibe, and Naito}}]{skinta2002}
\bibinfo{author}{\bibfnamefont{J.~A.} \bibnamefont{Skinta}},
  \bibinfo{author}{\bibfnamefont{M.-S.} \bibnamefont{Kim}},
  \bibinfo{author}{\bibfnamefont{T.~R.} \bibnamefont{Lemberger}},
  \bibinfo{author}{\bibfnamefont{T.}~\bibnamefont{Greibe}}, \bibnamefont{and}
  \bibinfo{author}{\bibfnamefont{M.}~\bibnamefont{Naito}},
  \bibinfo{journal}{Phys.\ Rev. Lett.} \textbf{\bibinfo{volume}{88}},
  \bibinfo{pages}{207005} (\bibinfo{year}{2002}).

\bibitem[{\citenamefont{Kim et~al.}(2003)\citenamefont{Kim, Skinta, Lemberger,
  Tsukada, and Naito}}]{kim2003}
\bibinfo{author}{\bibfnamefont{M.-S.} \bibnamefont{Kim}},
  \bibinfo{author}{\bibfnamefont{J.~A.} \bibnamefont{Skinta}},
  \bibinfo{author}{\bibfnamefont{T.~R.} \bibnamefont{Lemberger}},
  \bibinfo{author}{\bibfnamefont{A.}~\bibnamefont{Tsukada}}, \bibnamefont{and}
  \bibinfo{author}{\bibfnamefont{M.}~\bibnamefont{Naito}},
  \bibinfo{journal}{Phys. Rev. Lett.} \textbf{\bibinfo{volume}{91}},
  \bibinfo{pages}{087001} (\bibinfo{year}{2003}).

\bibitem[{\citenamefont{Biswas et~al.}(2002)\citenamefont{Biswas, Fournier,
  Qazilbash, Smolyaninova, Balci, and Greene}}]{biswas2002}
\bibinfo{author}{\bibfnamefont{A.}~\bibnamefont{Biswas}},
  \bibinfo{author}{\bibfnamefont{P.}~\bibnamefont{Fournier}},
  \bibinfo{author}{\bibfnamefont{M.~M.} \bibnamefont{Qazilbash}},
  \bibinfo{author}{\bibfnamefont{V.~N.} \bibnamefont{Smolyaninova}},
  \bibinfo{author}{\bibfnamefont{H.}~\bibnamefont{Balci}}, \bibnamefont{and}
  \bibinfo{author}{\bibfnamefont{R.~L.} \bibnamefont{Greene}},
  \bibinfo{journal}{Phys. Rev. Lett.} \textbf{\bibinfo{volume}{88}},
  \bibinfo{pages}{207004} (\bibinfo{year}{2002}).

\bibitem[{\citenamefont{Chesca et~al.}(2003)\citenamefont{Chesca, Ehrhardt,
  M\"ossle, Straub, Koelle, Kleiner, and Tsukada}}]{chesca2003}
\bibinfo{author}{\bibfnamefont{B.}~\bibnamefont{Chesca}},
  \bibinfo{author}{\bibfnamefont{K.}~\bibnamefont{Ehrhardt}},
  \bibinfo{author}{\bibfnamefont{M.}~\bibnamefont{M\"ossle}},
  \bibinfo{author}{\bibfnamefont{R.}~\bibnamefont{Straub}},
  \bibinfo{author}{\bibfnamefont{D.}~\bibnamefont{Koelle}},
  \bibinfo{author}{\bibfnamefont{R.}~\bibnamefont{Kleiner}}, \bibnamefont{and}
  \bibinfo{author}{\bibfnamefont{A.}~\bibnamefont{Tsukada}},
  \bibinfo{journal}{Phys. Rev. Lett.} \textbf{\bibinfo{volume}{90}},
  \bibinfo{pages}{057004} (\bibinfo{year}{2003}).

\bibitem[{\citenamefont{Chesca et~al.}(2005)\citenamefont{Chesca, Seifried,
  Dahm, Schopohl, Koelle, Kleiner, and Tsukada}}]{chesca2005}
\bibinfo{author}{\bibfnamefont{B.}~\bibnamefont{Chesca}},
  \bibinfo{author}{\bibfnamefont{M.}~\bibnamefont{Seifried}},
  \bibinfo{author}{\bibfnamefont{T.}~\bibnamefont{Dahm}},
  \bibinfo{author}{\bibfnamefont{N.}~\bibnamefont{Schopohl}},
  \bibinfo{author}{\bibfnamefont{D.}~\bibnamefont{Koelle}},
  \bibinfo{author}{\bibfnamefont{R.}~\bibnamefont{Kleiner}}, \bibnamefont{and}
  \bibinfo{author}{\bibfnamefont{A.}~\bibnamefont{Tsukada}},
  \bibinfo{journal}{Phys. Rev. B} \textbf{\bibinfo{volume}{71}},
  \bibinfo{pages}{104504} (\bibinfo{year}{2005}).

\bibitem[{\citenamefont{Shan et~al.}(2005)\citenamefont{Shan, Huang, Gao, Wang,
  Li, Dai, Zhou, Xiong, Ti, and Wen}}]{shan2005}
\bibinfo{author}{\bibfnamefont{L.}~\bibnamefont{Shan}},
  \bibinfo{author}{\bibfnamefont{Y.}~\bibnamefont{Huang}},
  \bibinfo{author}{\bibfnamefont{H.}~\bibnamefont{Gao}},
  \bibinfo{author}{\bibfnamefont{Y.}~\bibnamefont{Wang}},
  \bibinfo{author}{\bibfnamefont{S.~L.} \bibnamefont{Li}},
  \bibinfo{author}{\bibfnamefont{P.~C.} \bibnamefont{Dai}},
  \bibinfo{author}{\bibfnamefont{F.}~\bibnamefont{Zhou}},
  \bibinfo{author}{\bibfnamefont{J.~W.} \bibnamefont{Xiong}},
  \bibinfo{author}{\bibfnamefont{W.~X.} \bibnamefont{Ti}}, \bibnamefont{and}
  \bibinfo{author}{\bibfnamefont{H.~H.} \bibnamefont{Wen}},
  \bibinfo{journal}{Phys. Rev. B} \textbf{\bibinfo{volume}{72}},
  \bibinfo{pages}{144506} (\bibinfo{year}{2005}).

\bibitem[{\citenamefont{Balci and Greene}(2004)}]{balci2004}
\bibinfo{author}{\bibfnamefont{H.}~\bibnamefont{Balci}} \bibnamefont{and}
  \bibinfo{author}{\bibfnamefont{R.~L.} \bibnamefont{Greene}},
  \bibinfo{journal}{Phys. Rev. Lett.} \textbf{\bibinfo{volume}{93}},
  \bibinfo{pages}{067001} (\bibinfo{year}{2004}).

\bibitem[{\citenamefont{Stadlober et~al.}(1995)\citenamefont{Stadlober, Krug,
  Nemetschek, Hackl, Cobb, and Markert}}]{stadlober1995}
\bibinfo{author}{\bibfnamefont{B.}~\bibnamefont{Stadlober}},
  \bibinfo{author}{\bibfnamefont{G.}~\bibnamefont{Krug}},
  \bibinfo{author}{\bibfnamefont{R.}~\bibnamefont{Nemetschek}},
  \bibinfo{author}{\bibfnamefont{R.}~\bibnamefont{Hackl}},
  \bibinfo{author}{\bibfnamefont{J.~L.} \bibnamefont{Cobb}}, \bibnamefont{and}
  \bibinfo{author}{\bibfnamefont{J.~T.} \bibnamefont{Markert}},
  \bibinfo{journal}{Phys. Rev. Lett.} \textbf{\bibinfo{volume}{74}},
  \bibinfo{pages}{4911} (\bibinfo{year}{1995}).

\bibitem[{\citenamefont{Blumberg et~al.}(2002)\citenamefont{Blumberg, Koitzsch,
  Gozar, Dennis, Kendziora, Fournier, and Greene}}]{blumberg2002}
\bibinfo{author}{\bibfnamefont{G.}~\bibnamefont{Blumberg}},
  \bibinfo{author}{\bibfnamefont{A.}~\bibnamefont{Koitzsch}},
  \bibinfo{author}{\bibfnamefont{A.}~\bibnamefont{Gozar}},
  \bibinfo{author}{\bibfnamefont{B.~S.} \bibnamefont{Dennis}},
  \bibinfo{author}{\bibfnamefont{C.~A.} \bibnamefont{Kendziora}},
  \bibinfo{author}{\bibfnamefont{P.}~\bibnamefont{Fournier}}, \bibnamefont{and}
  \bibinfo{author}{\bibfnamefont{R.~L.} \bibnamefont{Greene}},
  \bibinfo{journal}{Phys. Rev. Lett.} \textbf{\bibinfo{volume}{88}},
  \bibinfo{pages}{107002} (\bibinfo{year}{2002}).

\bibitem[{qaz()}]{qazilbash2005}
\bibinfo{note}{{M. M. Qazilbash, B.S. Dennis, C. A. Kendziora, H. Balci, R. L.
  Greene, and G. Blumberg}, cond-mat/0501362; {M. M. Qazilbash, A. Koitzsch, B.
  S. Dennis, A. Gozar, H. Balci, C. A. Kendziora, R. L. Greene, and G.
  Blumberg}, cond-mat/0510098.}

\bibitem[{\citenamefont{Matsui et~al.}(2005{\natexlab{a}})\citenamefont{Matsui,
  Terashima, Sato, Takahashi, Fujita, and Yamada}}]{matsui2005a}
\bibinfo{author}{\bibfnamefont{H.}~\bibnamefont{Matsui}},
  \bibinfo{author}{\bibfnamefont{K.}~\bibnamefont{Terashima}},
  \bibinfo{author}{\bibfnamefont{T.}~\bibnamefont{Sato}},
  \bibinfo{author}{\bibfnamefont{T.}~\bibnamefont{Takahashi}},
  \bibinfo{author}{\bibfnamefont{M.}~\bibnamefont{Fujita}}, \bibnamefont{and}
  \bibinfo{author}{\bibfnamefont{K.}~\bibnamefont{Yamada}},
  \bibinfo{journal}{Phys. Rev. Lett.} \textbf{\bibinfo{volume}{95}},
  \bibinfo{pages}{017003} (\bibinfo{year}{2005}{\natexlab{a}}).

\bibitem[{\citenamefont{Ariando et~al.}(2005)\citenamefont{Ariando, Darminto,
  Smilde, Leca, Blank, Rogalla, and Hilgenkamp}}]{ariando2005}
\bibinfo{author}{\bibnamefont{Ariando}},
  \bibinfo{author}{\bibfnamefont{D.}~\bibnamefont{Darminto}},
  \bibinfo{author}{\bibfnamefont{H.~J.~H.} \bibnamefont{Smilde}},
  \bibinfo{author}{\bibfnamefont{V.}~\bibnamefont{Leca}},
  \bibinfo{author}{\bibfnamefont{D.~H.~A.} \bibnamefont{Blank}},
  \bibinfo{author}{\bibfnamefont{H.}~\bibnamefont{Rogalla}}, \bibnamefont{and}
  \bibinfo{author}{\bibfnamefont{H.}~\bibnamefont{Hilgenkamp}},
  \bibinfo{journal}{Phys. Rev. Lett.} \textbf{\bibinfo{volume}{94}},
  \bibinfo{pages}{167001} (\bibinfo{year}{2005}).

\bibitem[{\citenamefont{Devereaux and Einzel}(1995)}]{devereaux1995}
\bibinfo{author}{\bibfnamefont{T.~P.} \bibnamefont{Devereaux}}
  \bibnamefont{and} \bibinfo{author}{\bibfnamefont{D.}~\bibnamefont{Einzel}},
  \bibinfo{journal}{Phys. Rev. B} \textbf{\bibinfo{volume}{51}},
  \bibinfo{pages}{16336} (\bibinfo{year}{1995}).

\bibitem[{\citenamefont{Wang et~al.}(1991)\citenamefont{Wang, Chien, Ong,
  Tarascon, and Wang}}]{wang1991}
\bibinfo{author}{\bibfnamefont{Z.~Z.} \bibnamefont{Wang}},
  \bibinfo{author}{\bibfnamefont{T.~R.} \bibnamefont{Chien}},
  \bibinfo{author}{\bibfnamefont{N.~P.} \bibnamefont{Ong}},
  \bibinfo{author}{\bibfnamefont{J.~M.} \bibnamefont{Tarascon}},
  \bibnamefont{and} \bibinfo{author}{\bibfnamefont{E.}~\bibnamefont{Wang}},
  \bibinfo{journal}{Phys. Rev. B} \textbf{\bibinfo{volume}{43}},
  \bibinfo{pages}{3020} (\bibinfo{year}{1991}).

\bibitem[{\citenamefont{Jiang et~al.}(1994)\citenamefont{Jiang, Mao, Xi, Jiang,
  Peng, Venkatesan, Lobb, and Greene}}]{jiang1994}
\bibinfo{author}{\bibfnamefont{W.}~\bibnamefont{Jiang}},
  \bibinfo{author}{\bibfnamefont{S.~N.} \bibnamefont{Mao}},
  \bibinfo{author}{\bibfnamefont{X.~X.} \bibnamefont{Xi}},
  \bibinfo{author}{\bibfnamefont{X.}~\bibnamefont{Jiang}},
  \bibinfo{author}{\bibfnamefont{J.~L.} \bibnamefont{Peng}},
  \bibinfo{author}{\bibfnamefont{T.}~\bibnamefont{Venkatesan}},
  \bibinfo{author}{\bibfnamefont{C.~J.} \bibnamefont{Lobb}}, \bibnamefont{and}
  \bibinfo{author}{\bibfnamefont{R.~L.} \bibnamefont{Greene}},
  \bibinfo{journal}{Phys. Rev. Lett.} \textbf{\bibinfo{volume}{73}},
  \bibinfo{pages}{1291} (\bibinfo{year}{1994}).

\bibitem[{\citenamefont{Fournier et~al.}(1997)\citenamefont{Fournier, Jiang,
  Jiang, Mao, Venkatesan, Lobb, and Greene}}]{fournier1997}
\bibinfo{author}{\bibfnamefont{P.}~\bibnamefont{Fournier}},
  \bibinfo{author}{\bibfnamefont{X.}~\bibnamefont{Jiang}},
  \bibinfo{author}{\bibfnamefont{W.}~\bibnamefont{Jiang}},
  \bibinfo{author}{\bibfnamefont{S.~N.} \bibnamefont{Mao}},
  \bibinfo{author}{\bibfnamefont{T.}~\bibnamefont{Venkatesan}},
  \bibinfo{author}{\bibfnamefont{C.~J.} \bibnamefont{Lobb}}, \bibnamefont{and}
  \bibinfo{author}{\bibfnamefont{R.~L.} \bibnamefont{Greene}},
  \bibinfo{journal}{Phys. Rev. B} \textbf{\bibinfo{volume}{56}},
  \bibinfo{pages}{14149} (\bibinfo{year}{1997}).

\bibitem[{\citenamefont{Armitage et~al.}(2002)\citenamefont{Armitage, Ronning,
  Lu, Kim, Damascelli, Shen, Feng, Eisaki, Shen, Mang et~al.}}]{armitage2002}
\bibinfo{author}{\bibfnamefont{N.~P.} \bibnamefont{Armitage}},
  \bibinfo{author}{\bibfnamefont{F.}~\bibnamefont{Ronning}},
  \bibinfo{author}{\bibfnamefont{D.~H.} \bibnamefont{Lu}},
  \bibinfo{author}{\bibfnamefont{C.}~\bibnamefont{Kim}},
  \bibinfo{author}{\bibfnamefont{A.}~\bibnamefont{Damascelli}},
  \bibinfo{author}{\bibfnamefont{K.~M.} \bibnamefont{Shen}},
  \bibinfo{author}{\bibfnamefont{D.~L.} \bibnamefont{Feng}},
  \bibinfo{author}{\bibfnamefont{H.}~\bibnamefont{Eisaki}},
  \bibinfo{author}{\bibfnamefont{Z.~X.} \bibnamefont{Shen}},
  \bibinfo{author}{\bibfnamefont{P.~K.} \bibnamefont{Mang}},
  \bibnamefont{et~al.}, \bibinfo{journal}{Phys. Rev. Lett.}
  \textbf{\bibinfo{volume}{88}}, \bibinfo{pages}{257001}
  (\bibinfo{year}{2002}).

\bibitem[{\citenamefont{Matsui et~al.}(2005{\natexlab{b}})\citenamefont{Matsui,
  Terashima, Sato, Takahashi, Wang, Yang, Ding, Uefuji, and
  Yamada}}]{matsui2005b}
\bibinfo{author}{\bibfnamefont{H.}~\bibnamefont{Matsui}},
  \bibinfo{author}{\bibfnamefont{K.}~\bibnamefont{Terashima}},
  \bibinfo{author}{\bibfnamefont{T.}~\bibnamefont{Sato}},
  \bibinfo{author}{\bibfnamefont{T.}~\bibnamefont{Takahashi}},
  \bibinfo{author}{\bibfnamefont{S.-C.} \bibnamefont{Wang}},
  \bibinfo{author}{\bibfnamefont{H.-B.} \bibnamefont{Yang}},
  \bibinfo{author}{\bibfnamefont{H.}~\bibnamefont{Ding}},
  \bibinfo{author}{\bibfnamefont{T.}~\bibnamefont{Uefuji}}, \bibnamefont{and}
  \bibinfo{author}{\bibfnamefont{K.}~\bibnamefont{Yamada}},
  \bibinfo{journal}{Phys. Rev. Lett.} \textbf{\bibinfo{volume}{94}},
  \bibinfo{pages}{047005} (\bibinfo{year}{2005}{\natexlab{b}}).

\bibitem[{\citenamefont{Kusko et~al.}(2002)\citenamefont{Kusko, Markiewicz,
  Lindroos, and Bansil}}]{kusko2002}
\bibinfo{author}{\bibfnamefont{C.}~\bibnamefont{Kusko}},
  \bibinfo{author}{\bibfnamefont{R.~S.} \bibnamefont{Markiewicz}},
  \bibinfo{author}{\bibfnamefont{M.}~\bibnamefont{Lindroos}}, \bibnamefont{and}
  \bibinfo{author}{\bibfnamefont{A.}~\bibnamefont{Bansil}},
  \bibinfo{journal}{Phys. Rev. B} \textbf{\bibinfo{volume}{66}},
  \bibinfo{pages}{140513(R)} (\bibinfo{year}{2002}).

\bibitem[{\citenamefont{Yuan et~al.}(2004)\citenamefont{Yuan, Chen, Lee, and
  Ting}}]{yuan2004}
\bibinfo{author}{\bibfnamefont{Q.}~\bibnamefont{Yuan}},
  \bibinfo{author}{\bibfnamefont{Y.}~\bibnamefont{Chen}},
  \bibinfo{author}{\bibfnamefont{T.~K.} \bibnamefont{Lee}}, \bibnamefont{and}
  \bibinfo{author}{\bibfnamefont{C.~S.} \bibnamefont{Ting}},
  \bibinfo{journal}{Phys. Rev. B} \textbf{\bibinfo{volume}{69}},
  \bibinfo{pages}{214523} (\bibinfo{year}{2004}).

\bibitem[{\citenamefont{Voo and Wu}(2005)}]{voo2005}
\bibinfo{author}{\bibfnamefont{K.~K.} \bibnamefont{Voo}} \bibnamefont{and}
  \bibinfo{author}{\bibfnamefont{W.~C.} \bibnamefont{Wu}},
  \bibinfo{journal}{Physica C} \textbf{\bibinfo{volume}{417}},
  \bibinfo{pages}{103} (\bibinfo{year}{2005}).

\bibitem[{\citenamefont{Luo and Xiang}(2005)}]{luo2005}
\bibinfo{author}{\bibfnamefont{H.~G.} \bibnamefont{Luo}} \bibnamefont{and}
  \bibinfo{author}{\bibfnamefont{T.}~\bibnamefont{Xiang}},
  \bibinfo{journal}{Phys.\ Rev. Lett.} \textbf{\bibinfo{volume}{94}},
  \bibinfo{pages}{027001} (\bibinfo{year}{2005}).

\bibitem[{\citenamefont{Xiang and Wheatley}(1996)}]{Xiang1996}
\bibinfo{author}{\bibfnamefont{T.}~\bibnamefont{Xiang}} \bibnamefont{and}
  \bibinfo{author}{\bibfnamefont{J.~M.} \bibnamefont{Wheatley}},
  \bibinfo{journal}{Phys. Rev. Lett.} \textbf{\bibinfo{volume}{76}},
  \bibinfo{pages}{134} (\bibinfo{year}{1996}).

\bibitem[{csl()}]{csliu}
\bibinfo{note}{C. S. Liu, unpublished notes.}

\bibitem[{\citenamefont{{K. Yamada and K. Kurahashi and T. Uefuji and M. Fujita
  and S. Park and S.-H. Lee and Y. Endoh}}(2003)}]{yamada137004}
\bibinfo{author}{\bibnamefont{{K. Yamada and K. Kurahashi and T. Uefuji and M.
  Fujita and S. Park and S.-H. Lee and Y. Endoh}}}, \bibinfo{journal}{Phys.\
  Rev. Lett.} \textbf{\bibinfo{volume}{90}}, \bibinfo{pages}{137004}
  (\bibinfo{year}{2003}).

\end{thebibliography}

\end{document}